# Improving Company Valuations with Automated Knowledge Discovery, Extraction and Fusion[1]

Albert Weichselbraun, Philipp Kuntschik, Sandro Hörler
Swiss Institute for Information Science
University of Applied Sciences of the Grisons
Chur, Switzerland

**Abstract**

Performing company valuations within the domain of biotechnology, pharmacy and medical technology is a *challenging* task, especially when considering the unique set of risks biotech start-ups face when entering new markets. Companies specialized in global valuation services, therefore, combine valuation models and past experience with heterogeneous metrics and indicators that provide insights into a company's performance.

This paper illustrates how automated knowledge discovery, extraction and data fusion can be used to (i) obtain additional indicators that provide insights into the success of a company's product development efforts, and (ii) support labor-intensive data curation processes. We apply deep web knowledge acquisition methods to identify and harvest data on clinical trials that is hidden behind proprietary search interfaces and integrate the extracted data into the industry partner's company valuation ontology. In addition, focused Web crawls and shallow semantic parsing yield information on the company's key personnel and respective contact data, notifying domain experts of relevant changes that get then incorporated into the industry partner's company data.

## Introduction

Company valuations provide independent assessments which include information on the economic value of start-ups, their products and technologies. Performing such valuations necessitates comprehensive background information on the company's product-pipelines, financing, licensing contracts, management and success metrics. Gathering these data points requires significant resources and investments in the data acquisition and curation process.

The presented research was conducted together with a Swiss company specialized in company valuations and focuses on improving the efficiency and effectiveness of these data acquisition and curation processes by developing machine learning components that

1. automatically identify and retrieve structured data on a company's product pipeline (i.e. products and medications participating in clinical trials) from government clinical trial platforms such as *ClinicalTrials.gov* operated by the U.S. National Institutes of Health, the EU Clinical Trials Register and the World Health Organization's WHO clinical trials platform.

---

2. locate changes to a company's management and contact information communicated through its Web presence.

The knowledge retrieved by automated components is then integrated into the industry partner's business development databases and domain ontology.

**Related Work**

Deep web data acquisition develops strategies for obtaining web data that is hidden behind proprietary search interfaces. Research estimates that the deep web contains data in a scale even bigger than the content accessible through search engines which is, in contrast, referred to as surface web (Khelghati et al., 2013; Noor et al., 2011). A comprehensive study performed by (Chang et al., 2004) in 2004 estimated that the deep web contains over 450,000 web databases of which more than three quarters expose structured content over a query interface. (He et al., 2013) consider query generation, empty page filtering and URL deduplication important sub-problems of obtaining data from deep web sites and present techniques for addressing these challenges.

Named entity linking is a knowledge extraction technique used for grounding mentions of products, medications and companies to the corresponding entries in a target ontology. (Gangemi, 2013) provides an overview of knowledge extraction tools including specific applications for named entity recognition and linking. (C. Wang et al., 2012). approach named entity linking by suggesting a graph-based model (MentionRank), leveraging the principle that homogeneous groups of entities often occur in similar documents. This context-awareness helps to disambiguate terms such as "Apple" or "HP" when they occur in documents with an information technology or business focus. Recent approaches such as (Weichselbraun et al., 2019) use machine learning for various subtasks of the named entity linking process including generating name variances or performing the actual linking step. Shallow semantic parsing which is also known as slot filling draws upon the extracted named entities and assigns roles (e.g. chiefExecuteOfficerOf, phoneNumberOf, etc.) to them which allows combining multiple pieces of information (e.g. name, address, phone number, etc.) into a single knowledge base entry. Since shallow semantic parsing draws upon the results of multiple knowledge extraction techniques and even small inaccuracies multiply, it is particularly susceptible to mistakes. This problem is also reflected in the outcomes of semantic parsing competitions such as the *TAC 2017 Cold Start Slot Filling Task* in which even the winning system only obtained an F-measure below 20% (Lim et al., 2017). At the less complex *TAC KBP 2013 English Slot Filling Task* the top-ranked system developed by (Roth et al., 2014) yielded an F1 score of 37.3%. The system combined distant supervision with query-based relation extraction, therefore, modeling relation prediction as a classification task.

Nevertheless, it is important to keep in mind that the accuracy of shallow semantic parsing tasks can be significantly improved by reducing task complexity. (X. Wang et al., 2019), for instance, achieve an F1 score of over 76.5% for a temporal slot filling task that solely aims at determining a country's president from News articles. Similar to their efforts, the research presented in this paper, successfully applies shallow semantic parsing to isolated, well-defined knowledge extraction tasks required for providing company valuations of companies operating in the biotech domain.

**Method**

Significant parts of the World Wide Web are hard to grasp for search engines since they are composed of dynamic Web pages, generated from comprehensive topic-specific databases and only accessible through specific portal pages. These Deep Web data sources are often of particular importance to industrial applications, since they contain high quality, curated and up-to-date content on various subjects such as scientific papers, library catalogs and medical trials. The process outlined in this paper combines data acquisition techniques such as deep web mirroring and focused web crawling with knowledge extraction and data fusion methods to integrate information extracted from these sources with existing domain ontologies and databases.

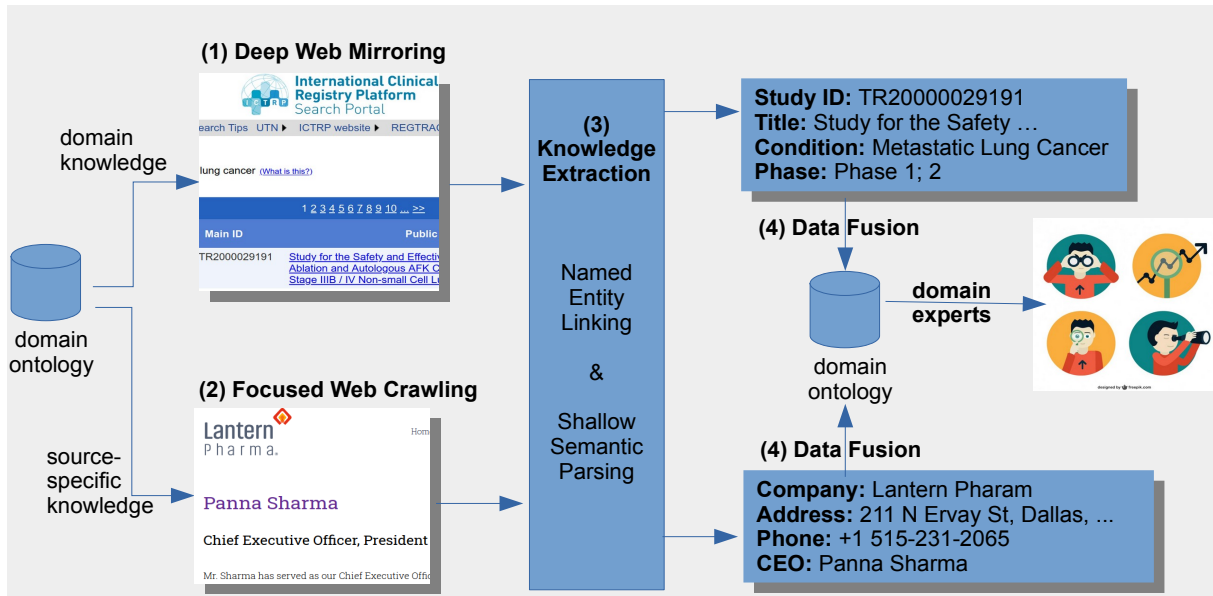

*Figure 1: Overview of the automated knowledge discovery, extraction and fusion process.*

Figure 1 illustrates the developed process that also considers background knowledge encoded in the industry partner's domain ontologies. The deep web mirroring component (1), for instance, uses this knowledge for optimizing the sequence in which data is queried from deep web resources. Once a clinical trial has been retrieved, we use the Recognyze named entity linking framework (Weichselbraun et al., 2015) in conjunction with shallow semantic parsing (3) for the knowledge extraction step which identifies information such as the companies and research institutions involved in the clinical trial, the study's title, approach, indications and progress. Afterwards, this information is normalized and fused with existing records (4) in the industry partner's domain ontology.

A second, parallel data acquisition pipeline uses focused Web crawls (2) to monitor company Web sites for changes in the management structure or contact information. Again we use knowledge extraction (3) to identify the company's key personnel and address data, and integrate the extracted records (4) into the industry partner's domain ontology.

Both shallow semantic parsing tasks benefit tremendously from the use of background knowledge, obtained from the industry partner's domain ontology and other domain-specific knowledge sources that have been created within the project which allowed

1. substituting named entity recognition with named entity linking which connects named entities (i.e. organizations, names of key personnel, etc.) to the corresponding knowledge base entries and has been proven to be much more accurate in the given setting;

2. mapping different specifications of the clinical phase within the deep web resources to a standardized values;

3. accurate identification of phone numbers since background information on their correct composition depending on country and region has been available to the information extraction components.

Limiting semantic parsing to very specific roles such as *clinicalPhaseOf*, *performedBy*, *isPhoneNumberOf*, etc. further improved the accuracy of the applied components.

**Results**

A prototype of the Deep Web data acquisition component yielded over 480,000 clinical trials from which 61,865 have been successfully linked to companies in the industry partner's domain ontology. A total of 35,757 clinical studies (7.4% of all retrieved trials) were flagged as completed at the time of this analysis.

The focused crawling component retrieved over 1.3 million web sites from 53,335 distinct companies validating over 233,000 records of key personnel operating in the biotech domain. These numbers clearly indicate that a manual curation process would be economically unviable and demonstrate the potential of the introduced methods in commercial settings.

The innovations developed within the project yield comprehensive and timely information on clinical trials and companies at very low cost, enabling the industry partner to base their valuations of biotech and pharma start-ups on an even more solid base. The clinical trial dataset has been integrated into the industry partner's domain ontology and also been made available to customers as part of an investor and business development database.

**Acknowledgement**

The DISCOVER project (https://www.fhgr.ch/discover) has been funded by Innosuisse. This support has been instrumental in facilitating the transfer of state-of-the art research methods to commercially settings within an innovative Swiss company.

**References**

1. Chang, K. C.-C., He, B., Li, C., Patel, M., & Zhang, Z. (2004). Structured Databases on the Web: Observations and Implications. *SIGMOD Rec.*, *33*(3), 61–70. https://doi.org/10.1145/1031570.1031584

2. Gangemi, A. (2013). A Comparison of Knowledge Extraction Tools for the Semantic Web. In P. Cimiano, O. Corcho, V. Presutti, L. Hollink, & S. Rudolph (Eds.), *The Semantic Web: Semantics and Big Data* (pp. 351–366). Springer Berlin Heidelberg. http://link.springer.com/chapter/10.1007/978-3-642-38288-8_24

3. He, Y., Xin, D., Ganti, V., Rajaraman, S., & Shah, N. (2013). Crawling Deep Web Entity Pages. *Proceedings of the Sixth ACM International Conference on Web Search and Data Mining*, 355–364. https://doi.org/10.1145/2433396.2433442

4. Khelghati, M., Hiemstra, D., & Van Keulen, M. (2013). Deep Web Entity Monitoring. *Proceedings of the 22Nd International Conference on World Wide Web*, 377–382. https://doi.org/10.1145/2487788.2487946

5. Lim, S., Kwon, S., Lee, S., & Choi, J. (2017). UNIST SAIL System for TAC 2017 Cold Start Slot Filling. *TAC*.

6. Noor, U., Rashid, Z., & Rauf, A. (2011). Article: A Survey of Automatic Deep Web Classification Techniques. *International Journal of Computer Applications*, *19*(6), 43–50.

7. Roth, B., Barth, T., Wiegand, M., Singh, M., & Klakow, D. (2014). Effective Slot Filling Based on Shallow Distant Supervision Methods. *ArXiv:1401.1158 [Cs]*. http://arxiv.org/abs/1401.1158

8. Wang, C., Chakrabarti, K., Cheng, T., & Chaudhuri, S. (2012). Targeted disambiguation of ad-hoc, homogeneous sets of named entities. *Proceedings of the 21st International Conference on World Wide Web*, 719–728. https://doi.org/10.1145/2187836.2187934

9. Wang, X., Zhang, H., Li, Q., Shi, Y., & Jiang, M. (2019). A Novel Unsupervised Approach for Precise Temporal Slot Filling from Incomplete and Noisy Temporal Contexts. *The World Wide Web Conference*, 3328–3334. https://doi.org/10.1145/3308558.3313435

10. Weichselbraun, A., Kuntschik, P., & Brasoveanu, A. M. P. (2019). *Name Variants for Improving Entity Discovery and Linking*. Second conference on Language, Data and Knowledge (LDK 2019), Leipzig, Germany.

11. Weichselbraun, A., Streiff, D., & Scharl, A. (2015). Consolidating Heterogeneous Enterprise Data for Named Entity Linking and Web Intelligence. *International Journal on Artificial Intelligence Tools*, *24*(2).